\documentclass[12pt]{JHEP3}
\usepackage[centertags]{amsmath}
\usepackage{amssymb}
\usepackage{epsfig} 
\usepackage{amsmath}
\usepackage{graphicx}

\newcommand{\beq}{\begin{equation}}
\newcommand{\eeq}{\end{equation}}
\newcommand{\bea}{\begin{eqnarray}}
\newcommand{\eea}{\end{eqnarray}}
\newcommand{\ba}{\begin{array}}
\newcommand{\ea}{\end{array}}
\newcommand{\bi}{\begin{itemize}}
\newcommand{\ei}{\end{itemize}}
\newcommand{\bn}{\begin{enumerate}}
\newcommand{\en}{\end{enumerate}}
\newcommand{\bc}{\begin{center}}
\newcommand{\ec}{\end{center}}
\renewcommand{\l}{\left}
\renewcommand{\r}{\right}

\newcommand{\ol}{\overline}

\newcommand{\Ga}{\Gamma}
\newcommand{\De}{\Delta}

\newcommand{\al}{\alpha}
\newcommand{\be}{\beta}

\newcommand{\ga}{\gamma}

\newcommand{\la}{\lambda}

\newcommand{\si}{\sigma}

\newcommand{\nl}{\nonumber\\}

\newcommand{\wt}[1]{\widetilde{#1}}
\newcommand{\ME}[1]{\langle{#1}\rangle}
\newcommand{\ket}[1]{|{#1}\rangle}
\def\btopik{{B} \to \pi K}
\def\bra#1{\left\langle #1\right|}
\def\ket#1{\left| #1\right\rangle}
\def\Bsdecay{\bar{B}_s \to J/\psi \phi}

\title{\boldmath Neutral Gauge Boson Contributions to the Dimuon
  Charge Asymmetry in $B$ Decays}

\author{
Ashutosh Kumar Alok$^a$, Seungwon Baek$^b$, David 
London$^a$
\\
$^a$ Physique des Particules, Universit\'e de Montr\'eal, \\ C.P. 6128,
succ. centre-ville, Montr\'eal, QC, Canada H3C 3J7 \\
$^b$ The Institute of Basic Science and Department of Physics,
Korea University, Seoul 136-701, Republic of Korea \\
E-mail: \email{alok@lps.umontreal.ca},
\email{sbaek@korea.ac.kr},
\email{london@lps.umontreal.ca}.
}

\preprint{UdeM-GPP-TH-10-193}

\abstract{ Recently, the D\O\ Collaboration measured the CP-violating
  like-sign dimuon charge asymmetry in neutral $B$ decays, finding a
  3.2$\sigma$ difference from the standard-model (SM) prediction.  A
  non-SM charge asymmetry $a_{\rm sl}^s$ suggests a new-physics (NP)
  contribution to $B_s$-$\ol{B}_s$ mixing.  In this case, in order to
  explain the measured value of $a_{\rm sl}^s$ within its 1$\si$
  range, NP must be present in $\Ga_{12}^s$, the absorptive part of
  the mixing. In this paper, we examine whether such an explanation is
  possible in models with flavor-changing $Z$ ($Z$FCNC) or $Z'$
  ($Z'$FCNC) gauge bosons. The models must also reproduce the measured
  values of the indirect CP asymmetry $S_{\psi\phi}$ in $B_s \to
  J/\psi \phi$, and $\De \Ga_s$, the $B_s$-$\ol{B}_s$ width
  difference.  We find that the $Z$FCNC model cannot reproduce the
  present measured values of $S_{\psi\phi}$ and $a_{\rm sl}^s$ within
  their 1$\sigma$ ranges.  On the other hand, in the $Z'$FCNC model,
  the values of all three observables can be simultaneously
  reproduced.  }


\begin{document}

\section{Introduction}

Flavor-changing neutral-current (FCNC) processes occur only at the
loop level in the standard model (SM), and are therefore a very
sensitive probe of physics beyond the SM.  Interestingly, recently
there have been several measurements of quantities in $B$ decays,
especially in the $b \to s$ transition, which differ from the
predictions of the SM. For example, i) the value of $\sin 2\be$
measured in several $b \to s$ penguin decays is found to be smaller
than that obtained in $B_d \to J/\psi K_S$~\cite{Lunghi}, ii) in
$\btopik$, the SM has some difficulty in accounting for all the
experimental measurements \cite{piK}, iii) the forward-backward
asymmetry in $B \to {K}^* \mu^+ \mu^-$ shows a small deviation from
the SM, especially at low momentum transfer squared~\cite{Alok}, iv)
the measurement of CP violation in $B_s \to J/\psi\phi$ suggests the
presence of a new-physics (NP) CP-violating phase in $B_s$-$\ol{B}_s$
mixing \cite{phi_s}. Although these effects are not significant enough
to claim NP, at least they indicate that flavor physics will still
play a very important role in the search for NP in the LHC era.  The
LHCb will provide much more precise data on the above-mentioned
observables.

Recently the D\O\ Collaboration reported an anomalously large
CP-violating like-sign dimuon charge asymmetry in the $B$ system
\cite{D0}. The asymmetry is
\bea
 A_{\rm sl}^b \equiv \frac{N_b^{++}  - N_b^{--}}{N_b^{++}  + N_b^{--}} ~,
\eea 
where $N_b^{\pm\pm}$ is the number of events of $b \ol{b} \to
\mu^{\pm} \mu^{\pm} X$. From the analysis of 6.1 fb$^{-1}$ of data,
D\O\ found
\bea
  A_{\rm sl}^b = -(9.57 \pm 2.51 \pm 1.46) \times 10^{-3} ~,
\label{eq:Asl}
\eea
where the first (second) error is statistical (systematic). This
exhibits a 3.2$\sigma$ deviation from the SM prediction, $A_{\rm
  sl}^{b,{\rm SM}} = \l( -2.3^{+0.5}_{-0.6} \r) \times
10^{-4}$~\cite{Lenz}.  The asymmetry can be written as a linear
combination of $B_d$ and $B_s$ contributions~\cite{D0_theory}
\bea
  A_{\rm sl}^b = (0.506 \pm 0.043) a_{\rm sl}^d  +  (0.494 \pm 0.043) a_{\rm sl}^s ~, 
\label{eq:A_a}
\eea
where the semileptonic ``wrong-charge'' asymmetry $a_{\rm sl}^q$ $(q=d,s)$ is given by
\bea
  a_{\rm sl}^q = \frac{\Ga(\ol{B}_q \to \mu^+ X)-\Ga({B}_q \to \mu^- X)}
{\Ga(\ol{B}_q \to \mu^+ X)+\Ga({B}_q \to \mu^- X)} ~.
\eea
Thus, the discrepancy may indicate a NP contribution to $B_d$-$\ol{B}_d$
mixing and/or $B_s$-$\ol{B}_s$ mixing~\cite{D0_theory}.  

Using the current experimental value of $a_{\rm sl}^{d, {\rm exp}} =
-0.0047 \pm 0.0046$~\cite{D0,HFAG} and the SM prediction $a_{\rm
  sl}^{d, {\rm SM}} = (-4.8^{+1.0}_{-1.2}) \times 10^{-4}$,
Eq.~(\ref{eq:A_a}) leads to $a_{\rm sl}^s = -0.0146 \pm 0.0075$.  The
CDF Collaboration also measured $A_{\rm sl}^b$, but with much larger
errors: $A_{\rm sl}^b = (8.0 \pm 9.0 \pm 6.8) \times
10^{-3}$~\cite{CDF}.  Finally, D\O\ directly measured $a_{\rm sl}^s$:
$a_{\rm sl}^{s, {\rm exp}} = -(1.7 \pm 9.1^{+1.4}_{-1.5}) \times
10^{-3}$~\cite{D0_s}.  Combining all these results, we get
\bea
  (a_{\rm sl}^{s})_{\rm ave} = -(12.7 \pm 5.0) \times 10^{-3} ~,
\label{eq:asl}
\eea
which is still about 2.5$\sigma$ away from the SM prediction 
of $a_{\rm sl}^{s,{\rm SM}}=(2.1 \pm 0.6) \times 10^{-5}$.
 
The D\O\ measurement thus suggests the presence of NP in
$B_s$-$\ol{B}_s$ mixing, and we explore this possibility below. We
begin with a general review of the mixing.  In the $B_s$ system, the
mass eigenstates $B_{L}$ and $B_{H}$ ($L$ and $H$ indicate the light
and heavy states, respectively) are admixtures of the flavor
eigenstates $B_s$ and $\ol{B}_s$:
\bea 
\ket{B_L} &=& p \ket{B_s} + q \ket{\ol{B}_s} ~, \nl
\ket{B_H} &=& p \ket{B_s} - q \ket{\ol{B}_s} ~, 
\eea 
with $|p|^2 + |q|^2 =1$.  As a result, the initial flavor eigenstates
oscillate into one another according to the Schr\"odinger equation
\bea 
i \frac{d}{dt} \l( \ba{c} \ket{B_s(t)} \\ \ket{\ol{B}_s(t)} \ea \r) =
\l(M^s -i \frac{\Ga^s}{2} \r) \l( \ba{c} \ket{B_s(t)}
\\ \ket{\ol{B}_s(t)} \ea \r) ~,
\eea 
where $M=M^\dagger$ and $\Ga=\Ga^\dagger$ correspond respectively to
the dispersive and absorptive parts of the mass matrix. The
off-diagonal elements, $M^s_{12}=M_{21}^{s*}$ and
$\Ga^s_{12}=\Ga_{21}^{s*}$, are generated by $B_s$-$\ol{B}_s$
mixing. We define
\bea 
\Ga_s \equiv \frac{\Ga_H
  + \Ga_L}{2}, \quad \De M_s \equiv   M_H - M_L, \quad \De \Ga_s \equiv   \Ga_L -
\Ga_H ~.
\eea 
Expanding the mass eigenstates and $q/p$ in 
$\Ga_{12}^s/M_{12}^s$, we find, to a very good
approximation~\cite{Dunietz:2000}, 
\bea 
\De M_s &=& 2 |M_{12}^s| ~, \nl
\De \Ga_s &=& 2 |\Ga_{12}^s| \cos\phi_s  ~, \nl
{q
  \over p} &=& - e^{- i \phi_M^s} \l(1 -\frac{|M_{12}^s|}{2
  |\Ga_{12}^s|} \sin\phi_s \r) ~, 
\label{eq:mix_para}
\eea 
where $\phi_M^s \equiv \arg M_{12}^s$ and $\phi_s \equiv
\arg(-M_{12}^s/\Ga_{12}^s)$. Then $a_{\rm sl}^s$ is given by~\cite{Lenz}
\bea
a_{\rm sl}^s = {\rm Im}\l(\Ga_{12}^s \over M_{12}^s\r) =
\frac{|\Ga_{12}^{s}|}{|M_{12}^{s}|} \, \sin\phi_s ~.
\eea

In the SM, we have
\bea
a_{\rm sl}^s = (4.97 \pm 0.94) \times 10^{-3} \, \sin\phi_s ~.
\eea
Within the SM, the phase of $M_{12}^s$ is given by $\arg[(V_{tb}
  V_{ts}^*)^2]$ and that of $\Ga_{12}^s$ is dominated by $\arg[(V_{cb}
  V_{cs}^*)^2]$.  The state-of-the-art calculation gives $\phi_s =
0.0041 \pm 0.0008$~\cite{Lenz}. As noted above, the SM cannot explain
the D\O\ result; this is due to its far-too-small weak phases. From
here on, we neglect the SM weak phases.

Suppose now that NP contributes to $M_{12}^s$ but not to
$\Ga_{12}^s$. We introduce a new parameter $\De_s$ to take this into
account: $M_{12}^s = M_{12}^{s,{\rm SM}} + M_{12}^{s,{\rm NP}} =
M_{12}^{s,{\rm SM}} \De_s = M_{12}^{s,{\rm SM}} |\De_s| e^{i
  \phi_M^s}$.  We have
\bea
a_{\rm sl}^s = \frac{|\Ga_{12}^{s}|}{|M_{12}^{s,{\rm SM}}|} \frac{\sin\phi_M^s}{|\De_s|}=
(4.97 \pm 0.94) \times 10^{-3} \, \frac{\sin\phi_M^s}{|\De_s|} ~.
\label{eq:asl_rel}
\eea
The experimental value of the mass difference in the $B_s$-$\ol{B}_s$
system, $\De M_s = (17.77 \pm 0.12) \; {\rm ps}^{-1} $, constrains
$|\De_s| = 0.92 \pm 0.32$.  Then Eqs.~(\ref{eq:asl}) and
(\ref{eq:asl_rel}) give $\sin \phi_M^s = -2.56 \pm 1.16$, implying
that the full 1$\si$ range of the experimental result lies outside the
physical region. It is therefore not possible to explain the 1$\si$
range of the D\O\ measurement if NP contributes only to $M_{12}^s$.

This problem can be solved if the new physics contributes to
$\Ga_{12}^s$. Here we introduce a second new parameter $\Xi_s$:
$\Ga_{12}^s = \Ga_{12}^{s,{\rm SM}} + \Ga_{12}^{s,{\rm NP}} =
\Ga_{12}^{s,{\rm SM}} \Xi_s = \Ga_{12}^{s,{\rm SM}} |\Xi_s| e^{i
  \phi_\Ga^s}$. We then have
\bea
 a_{\rm sl}^s &=& \frac{|\Ga_{12}^{s,{\rm SM}}|}{|M_{12}^{s,{\rm SM}}|}  \frac{|\Xi_s|}{|\De_s|}
   \sin(\phi_M^s-\phi_\Ga^s) \nl
 &=& (4.97 \pm 0.94) \times 10^{-3} \, \frac{|\Xi_s|}{|\De_s|}
    \sin(\phi_M^s-\phi_\Ga^s) ~.
\eea
Now Eq.~(\ref{eq:asl}) can be reproduced if $|\Xi_s|$ is sufficiently
large. 

The difficulty is that, in many NP models, the contribution to
$\Ga_{12}^s$ is not large enough to compete with that of the SM, which
is dominated by the tree-level $b \to s c \bar{c}$ process.  Two
exceptions are R-parity-violating supersymmetric models~\cite{D0_He}
and leptoquark models~\cite{leptoquark} (other analyses of
Eq.~(\ref{eq:asl}) in various NP models can be found in
Refs.~\cite{D0_theory,D0_theory2}).

In this paper we examine the contribution of flavor-changing neutral
gauge bosons to $\Ga_{12}^s$. We consider two types of models. They
involve tree-level $Z{\ol b}s$ or $Z'{\ol b}s$ couplings. As we will
see, the non-universal $Z'$ model can enhance $\Ga_{12}^s$ enough to
explain the 1$\si$ range of Eq.~(\ref{eq:asl}).

The paper is organised as follows. The NP contribution to $\Ga_{12}^s$
is discussed in Sec.~\ref{sec:Gamma12}. In Sec.~\ref{sec:model_dep},
we describe the models with flavor-changing $Z$ and $Z'$
couplings. The numerical results are presented in
Sec.~\ref{sec:num_res}. We conclude in Sec.~\ref{sec:concl}.

\section{\boldmath Contributions to $\Ga_{12}^s$}
\label{sec:Gamma12}

In the SM, the dominant contribution to $\Ga_{12}^{s,{\rm SM}}$ comes
from the charged-current $b \to s c \bar{c}$ operator with a $c$-quark
loop. It is given by
\bea
\Ga_{12}^{s,{\rm SM}}&=& -\frac{ G_F^2 m_b^2 \la_c^2}{3 \pi} \sqrt{1-4
  x_c} \times \nl 
&& \hskip-.7truecm \Bigg\{ \l[ K_1 (1-x_c) +{1 \over 2} K_2 (1-4
  x_c) \r] \ME{O_{LL}} + ( K_1 - K_2) (1+2x_c) \ME{\wt{O}_{RR}} \Bigg
\}, 
\label{eq:G12_SM}
\eea 
where $\la_c = V_{cb} V_{cs}^*$, $x_c = m_c^2/m_b^2$, and $K_1 = 3
C_1^2 + 2 C_1 C_2$, $K_2 = C_2^2$.  The values for the $C_i$ are
$C_1(m_b) = 1.086, C_2(m_b) = -0.197$ (Tables 4, 5 of
Ref.~\cite{BBL}).  We use the vacuum insertion approximation to
calculate the hadronic matrix elements~\cite{Becirevic}:
\bea
\ME{O_{LL}} \equiv \ME{B_s | \ol{s} \ga_\mu P_L b \; \ol{s} \ga^\mu P_L b  | \ol{B}_s} &=& {1 \over 3} m_{B_s} f_{B_s}^2 ~, \nl
\ME{\wt{O}_{RR}} \equiv \ME{B_s | \ol{s}  P_R b \; \ol{s} P_R b | \ol{B}_s} &=& -{5 \over 24} m_{B_s} f_{B_s}^2 \l(m_{B_s} \over m_b + m_s\r)^2 ~,
\label{vac}
\eea
where $P_{L,R} = (1\mp \gamma_5)/2$ and $f_{B_s}=(238.8 \pm 9.5)$ MeV
\cite{Laiho:2009eu}.  An updated theoretical prediction can be found
in Ref.~\cite{Lenz}, whose results we use in our numerical
calculations.

Any NP contribution to $\Ga_{12}^{s,{\rm NP}}$ must come from a new
operator of the form $b \to s f \ol{f}$, where $f$ is a light
fermion. If $\Ga_{12}^{s,{\rm NP}}$ is to be significant, it must be
at least comparable to $\Ga_{12}^{s,{\rm SM}}$. Now, most NP light
fermionic operators are constrained to be small. In particular, (i)
$e$ and $\mu$ loop contributions are strongly bounded by the $b \to s
e^+ e^-$ and $b \to s \mu^+ \mu^-$ processes, (ii) the $b \to s u
\ol{u}$ and $b \to s d \ol{d}$ operators are constrained by the
measurement of $B(\bar{B} \to \pi K)$, (iii) $B(\bar{B} \to \phi K)$ constrains $b
\to s s \ol{s}$. On the other hand, the bounds on the NP $b \to s
\tau^+ \tau^-$ transition are very weak. For example, the present
upper bound on $B(\bar{B}_s\to \tau^+\tau^-)$ is only $\lesssim 5
\%$~\cite{Grossman:1996}, to be compared with the SM prediction of
$B(\bar{B}_s\to \tau^+ \tau^-) \sim 10^{-8}$. Also, the current upper bound
on $B(\bar{B} \to X_s\, \tau^+ \,\tau^-)$ is just $\lesssim 5 \%$
\cite{Grossman:1996}.  For this reason, in the NP models we consider,
we examine the contribution to $\Ga_{12}^{s,{\rm NP}}$ coming from $b
\to s \tau^+ \tau^-$.

Another potential important effect on $\Ga_{12}^{s,{\rm NP}}$ comes
from the NP $b \to s c \bar{c}$ operator. Although the constraints from
$\bar{B} \to D D_s$, $\bar{B}_d \to J/\psi K_S$, etc.\ are such that the NP
$c$-quark loop contribution cannot be large enough to compete with
$\Ga_{12}^{s,{\rm SM}}$, the SM-NP interference term can be
significant.

The decay width difference $\De \Ga_s$ has been measured
independently.  The angular analysis of $\bar{B}_s \to J/\psi \phi$ gives
\cite{HFAG,Dighe:1995pd,Dighe:1998vk}
\beq
 \De \Ga_s = \pm (0.154^{+0.054}_{-0.070}) \; {\rm ps}^{-1} ~, 
\label{eq:DGamma_s}
\eeq
to be compared with the the SM prediction \cite{Lenz}
\beq
 \De \Ga_s^{\rm SM} = (0.096 \pm 0.039) \; {\rm ps}^{-1}~.
\label{eq:DGamma_s_SM}
\eeq
If NP contributes to $\Ga_{12}^{s,{\rm NP}}$, it is present in the
width difference, whose expression is given by
\beq
\De \Ga_s = \De \Ga_s^{\rm SM} \, |\Xi_s| \cos(\phi_M^s-\phi_\Ga^s)\;.
\eeq
The measurement of $\De \Ga_s$ therefore constrains $\Ga_{12}^{s,{\rm
    NP}}$.

There is another measurement which must be taken into account. The CDF
\cite{CDFpsiphi} and D\O\ \cite{D0psiphi} Collaborations have measured
indirect CP violation in $\bar{B}_s \to J/\psi \phi$. They obtain
$S_{\psi\phi} = -2\beta_s$, and find \cite{HFAG}
\beq
\beta_s = 0.39^{+0.18}_{-0.14} ~~{\rm or}~~ 1.18^{+0.14}_{-0.18} ~.
\eeq
This disagrees with the SM prediction
\beq
\beta_s^{\rm SM} = 0.019 \pm 0.001
\label{betasSM}
\eeq
at $2\sigma$. (Note: the recent CDF measurement agrees with the SM
better than the previous D\O and CDF measurements.)  Now, we have
assumed that the NP affects $M_{12}^s$, so that there is a weak phase
in $B_s$-$\ol{B}_s$ mixing, $\phi_M^s$. Depending on which NP
operator(s) contribute to $\Ga_{12}^{s,{\rm NP}}$, there may also be a
contribution to the decay of $\bar{B}_s \to J/\psi \phi$. Thus, the NP is
constrained in that the CDF/D\O\ measurement must be reproduced.

To detail the contribution of the NP to the indirect CP asymmetry in
$\bar{B}_s \to J/\psi \phi$, we follow the procedure of
Ref.~\cite{psiphigeneral}. First, we note that $\bar{B}_s \to J/\psi \phi$
is really three separate decays, one for each polarization state
$\lambda$ of the final-state vector particles; longitudinal: $\lambda
= 0$, transverse: $\lambda = \left\{\|,\perp \right\}$. Second, in
Ref.~\cite{DLNP} it is argued that all strong phases associated with
NP amplitudes are negligible. In this case, for each polarization one
can combine all NP matrix elements into a single NP amplitude, with a
single weak phase $\varphi_\lambda$:
\beq
\sum \bra{(J/\psi \phi)_\lambda} {\cal O}_{\rm NP} \ket{B_s}
= b_\lambda e^{i\varphi_\lambda} ~.
\eeq
We now assume that this single NP amplitude contributes to the decay
of $\Bsdecay$.  The decay amplitude for each of the three possible
polarization states may then be written as
\bea
A_\lambda \equiv Amp (B_s \to J/\psi \phi)_\lambda &=&
a_\lambda e^{i (\delta_\lambda^a -
\delta_\perp^a)} + b_\lambda e^{i\varphi_\lambda} e^{- i
\delta_\perp^a} ~, \nl
{\bar A}_\lambda \equiv Amp ({\bar B}_s \to J/\psi \phi)_\lambda &=&
a_\lambda e^{i (\delta_\lambda^a -
\delta_\perp^a)} + b_\lambda e^{-i\varphi_\lambda} e^{- i
\delta_\perp^a} ~,
\label{amps}
\eea
where $a_\lambda$ and $b_\lambda$ represent the SM and NP amplitudes,
respectively, $\varphi_\lambda$ is the new-physics weak phase, and the
$\delta_\lambda^a$ are the SM strong phases.  All strong phases are
given relative to $\delta_\perp^a$. $a_\lambda$ is defined to be
positive for every polarization. $b_\lambda$ can also be taken to be
positive: if it is negative, the minus sign can be absorbed in the
weak phase by redefining $\varphi_\lambda \to \varphi_\lambda +
\pi$. We emphasize this fact by writing the ratio
$b_\lambda/a_\lambda$ as the positive-definite quantity $|r_\lambda|$.
Note that strong phases are generated by rescattering, and this
costs a factor of about 25. The strong phase of the SM
color-suppressed ${\bar b} \to {\bar c} c {\bar s}$ diagram $C$ is
generated by rescattering of the color-allowed ${\bar b} \to {\bar c}
c {\bar s}$ tree diagram $T$.  Since $|C/T|$ is expected to be in the
range 0.2-0.6, the SM strong phase is on the small side, but is not
negligible.

Putting all this together, we find that the indirect CP asymmetry in
$\Bsdecay$ measures
\beq
S_{\psi\phi} = \sin\phi_M^s + 2 |r_\lambda| \cos\phi_M^s
 \sin\varphi_\lambda\cos\delta_{\lambda}^a ~.
\label{Spsiphi}
\eeq
(We have neglected the SM contribution to the weak phase of
$B_s$-$\ol{B}_s$ mixing, which is expected to be only $\sim 2\%$
[Eq.~(\ref{betasSM})].)  If the NP contribution to $\Ga_{12}^{s,{\rm
    NP}}$ comes only from $b \to s \tau^+ \tau^-$, then $r_\lambda =
0$ and $S_{\psi\phi} = \sin\phi_M^s$.  However, if the NP $b \to s c
\ol{c}$ operator is involved in $\Ga_{12}^{s,{\rm NP}}$, then
$r_\lambda \ne 0$ and the full expression for $S_{\psi\phi}$ above
must be used.

Finally, we note that the phases of $a_{\rm sl}^s$ and $S_{\psi\phi}$
are different when there is a NP contribution to $\Ga_{12}^s$ and/or
${\bar B}_s \to J/\psi \phi$.  In addition, the relation between
them~\cite{Grossman:2009},
\bea
a_{\rm sl}^s = -\frac{|\De \Ga_s|}{|\De M_s|} S_{\psi\phi}/\sqrt{1-S_{\psi\phi}^2} ~,
\eea
is violated in this case.

\section{New-Physics Models}
\label{sec:model_dep}

\subsection{\bf\boldmath $Z$-mediated FCNC's}

In the model with $Z$-mediated FCNC's ($Z$FCNC), a new vector-like
isosinglet down-type quark $d'$ is added to the particle spectrum
\cite{Grossman:1997, Barenboim:2001, Barenboim:1997pf}.  Such quarks
appear in $E_6$ GUT theories, for example. The ordinary $Q_{em} =
-1/3$ quarks mix with the $d'$.  Because the $d'_L$ has a different
$I_{3L}$ from $d_L$, $s_L$ and $b_L$, FCNC's appear at tree level in
the left-handed sector.  In particular, a $Z{\bar b}s$ coupling can be
generated:
\beq
{\cal L}^{ Z}_{ FCNC} = -\frac{g}{2 \cos\theta_{
W}} U_{sb} \, \bar s \gamma^\mu P_L b \, Z_\mu
+ {\rm h.c.} 
\label{Usb} 
\eeq
This coupling leads to a NP contribution to $B_s$-$\ol{B}_s$ mixing at
tree level. 

The SM contribution to $M_{12}^s$ is
\bea
M_{12}^{s,{\rm SM}} = \frac{G_F^2}{12 \pi^2} M_W^2 \la_t^2
 \eta_ B B_{B_s} f_{B_s}^2 M_{B_s} E(x_t) ~,
\eea
where $\la_t = V_{tb} V_{ts}^*$, $x_t = m_t^2/M_W^2$, $\eta_B\simeq
0.551$ is the QCD correction, and we take the hadronic parameter
$f_{B_s} B_{B_s}^{1/2} = 295 \pm 36$ MeV.  The loop function
$E(x_t)$ is given by
\bea
  E(x_t) = \frac{-4 x_t + 11 x_t^2- x_t^3  }{4(1-x_t)^2}
+\frac{3 x_t^3 \ln x_t}{2(1-x_t)^3} ~.
\eea

The mass difference $\De M_s$ in the $Z$FCNC model is given
by~\cite{Barenboim:1997pf}
\bea
 \De M_s = \De M_s^{\rm SM} \l| 
 1 + a \l(U_{sb} \over \la_t \r) - b \l(U_{sb} \over \la_t \r)^2
\r| ~,
\label{DMsZFCNC}
\eea
where
\bea
  a = 4 \frac{C(x_t)}{E(x_t)} ~~,~~~~
  b = \frac{2 \sqrt{2} \pi^2}{G_F M_W^2 E(x_t)} ~.
\eea
The loop functions $E(x_t)$ and $C(x_t)$ are given by~\cite{Barenboim:1997pf}
\bea
  E(x_t) & = & \frac{-4 x_t + 11 x_t^2- x_t^3  }{4(1-x_t)^2}
+\frac{3 x_t^3 \ln x_t}{2(1-x_t)^3}~,\nl
  C(x_t) & = & \frac{x_t}{4} \l[\frac{4-x_t}{1-x_t} 
+ \frac{3 x_t \ln x_t  }{(1-x_t)^2} \r] ~.
\eea
The term in Eq.~(\ref{DMsZFCNC}) proportional to $a$ is obtained from
a diagram with both SM and NP $Z$ vertices; that proportional to $b$
corresponds to the diagram with two NP $Z$ vertices.

This coupling will also lead to contributions to $\Ga_{12}^s$ due to
the generation of the $Z$-mediated operator $b \to s f \ol{f}$
($f=\tau$, $c$). The amplitude is
\beq
\frac{G_F}{\sqrt{2}} \, U_{sb} \, \bar s \gamma_\mu (1-\gamma_5)
b \, \bar f \gamma^\mu (I_3 - Q \sin^2\theta_{ W}) (1\mp
\gamma_5) f ~.
\eeq
There are diagrams with two NP vertices, and with one SM and one NP
vertex. The dominant contribution is due to SM-NP interference, yielding
\bea
\Ga_{12}^{s,Z} &=& -\frac{G_F^2\, m_b^2 \, \la_c\, U_{sb}}{\pi}
\sqrt{1-4 x_c} \times  \Bigg[ \l\{  {1 \over 2}(1-x_c) -{2 \over 3}
 \sin^2\theta_W(1+2x_c)\r\} \ME{O_{LL}} \nl 
&&
~~~~~~~~~~~ +~\l({1 \over 2} -{2 \over 3} \sin^2\theta_W \r) (1+2 x_c) \ME{\wt{O}_{RR}}
 \Bigg].
\eea

\subsection{\bf\boldmath $Z'$-mediated FCNC's}

In the model with $Z'$-mediated FCNC's ($Z'$FCNC), the gauge group
contains an additional $U(1)'$, which leads to a $Z'$
\cite{Langacker}. Within some string-construction or GUT models such
as $E_6$, it is possible to have family non-universal $Z'$
couplings. In the physical basis, FCNC's generally appear at tree
level in both the left-handed (LH) and right-handed (RH) sectors. In
particular, the interaction Lagrangian can contain $Z'{\bar b}s$
couplings:
\beq
{\cal L}^{ Z'}_{ FCNC} = -{g \over \cos\theta_W} \bigg[ \ol{s} \ga^\mu P_L B^L_{sb} b
 + (L \leftrightarrow R) \bigg] Z'_\mu + {\rm h.c.} ~,
\eeq
which lead to a tree-level contribution to $B_s$-$\ol{B}_s$
mixing. (Note: the $U(1)'$ coupling constant is $g'$. However, the
above Lagrangian is written in terms of the SM coupling; the ratio of
the two couplings is absorbed into the $B_{sb}$'s. This makes the
comparison of NP and SM effects more straightforward.)

The effective Hamiltonian for $M_{12}^s$ in the $Z'$FCNC model at the
scale $m_b$ is calculated from the tree-level diagrams
\cite{He:2006,Baek}:
\bea
{\cal H}_{\rm eff}&=& {4 G_F \over \sqrt{2}} {m_Z^2 \over m_{Z'}^2}
 \Bigg[ \eta^{6/23} (B^L_{sb})^2 O_{LL}(m_b) + \eta^{6/23} (B^R_{sb})^2  O_{RR}(m_b) 
+2 \eta^{3/23}  B^L_{sb} B^R_{sb} O_{LR}(m_b) \nl
&& ~~~~~~~~~~~~~~ +~{4 \over 3} \l(\eta^{3/23}-\eta^{-24/23}\r)  B^L_{sb} B^R_{sb} \wt{O}_{LR}(m_b)
 \Bigg] ~,
\label{effH-M12s}
\eea
where $\eta=\al_s(M_{Z'})/\al_s(m_b)$ and
\bea
O_{LL} &=& \ol{s} \ga_\mu P_L b \; \ol{s} \ga^\mu P_L b ~, \nl
O_{RR} &=& \ol{s} \ga_\mu P_R b \; \ol{s} \ga^\mu P_R b ~, \nl
O_{LR} &=& \ol{s} \ga_\mu P_L b \; \ol{s} \ga^\mu P_R b ~, \nl
 \wt{O}_{LR} &=& \ol{s} P_L b \;  \ol{s} P_R b ~.
\eea
In addition to the SM operator $O_{LL}$, the operators $O_{RR}$ and
$O_{LR}$ are generated at the scale $M_{Z'}$.  The operator
$\wt{O}_{LR}$ is generated through renormalization down to the scale
$m_b$.

The matrix element $M_{12}^{s,Z'}$ for $B_s$-$\ol{B}_s$ mixing is given by
\bea
M_{12}^{s,Z'} = \ME{B_s | {\cal H}_{\rm eff} | \ol{B}_s} ~.
\label{ME-M12s}
\eea
The hadronic matrix elements are again calculated using the vacuum
insertion approximation \cite{Becirevic}. $\ME{B_s | O_{LL} |
  \ol{B}_s}$ is given in Eq.~(\ref{vac}), $\ME{O_{RR}} \equiv \ME{B_s | O_{RR} |
  \ol{B}_s}=\ME{B_s | O_{LL} | \ol{B}_s}$, and
\bea
\ME{O_{LR}} \equiv \ME{B_s | O_{LR} | \ol{B}_s} &=& -\Big[{1 \over 4} + {1 \over 6} \l(m_{B_s} \over m_b + m_s\r)^2 \Big] m_{B_s} f_{B_s}^2 ~, \nl
\ME{\wt{O}_{LR}} \equiv \ME{B_s | \wt{O}_{LR} | \ol{B}_s} &=&  \Big[{1 \over 24} + {1 \over 4} \l(m_{B_s} \over m_b + m_s\r)^2 \Big] m_{B_s} f_{B_s}^2 ~.
\eea

The $Z'$ contribution to $\Ga_{12}^s$ through $\tau$ pairs is obtained
from the formulae given in Refs.~\cite{Golowich:2006,Chen:2007}:
\bea
\Ga_{12}^{s,Z'} &=& -\frac{ 4G_F^2 m_b^2}{3 \pi} \l(m_Z^2 \over m_{Z'}^2 \r)^2
 \sqrt{1-4 x_\tau} \times \nl
&& \hskip-2truemm \Bigg[ \Big\{ (B^{LL})^2 + (B^{RR})^2 + (B^{LR})^2 + (B^{RL})^2 \Big\}
 \Big\{ (1-x_\tau) \ME{O_{LL}} + (1+ 2 x_\tau) \ME{\wt{O}_{RR}} \Big\} \nl
&& +~ 2 \, \Big\{ B^{LL} B^{RL}+  B^{RR} B^{LR}\Big\}
 \Big\{ (1-x_\tau) \ME{O_{LR}} + (1+ 2 x_\tau) \ME{\wt{O}_{LR}} \Big\} \nl
&& +~ 6 \Big\{ B^{LL} B^{LR} +  B^{RR} B^{RL}\Big \} x_\tau \ME{O_{LL}} \nl 
&& +~ 6 \Big\{ B^{LL} B^{RR} +  B^{LR} B^{RL}\Big \} x_\tau \ME{O_{LR}} \Bigg],
\label{eq:G12_Zprime}
\eea
where $B^{ij}=B_{sb}^i B_{\tau\tau}^j~(i,j=L,R)$, $x_\tau =
m_\tau^2/m_b^2$.  Comparing Eqs.~(\ref{eq:G12_SM}) and
(\ref{eq:G12_Zprime}), we can see that if $B^{ij} \sim (\la_c
m_{Z'}^2/m_Z^2)^2 \sim {\cal O}(1)$, $\Ga_{12}^{s,Z'}$ can be
comparable to $\Ga_{12}^{s,{\rm SM}}$.  We will see that these ${\cal
  O}(1)$ couplings are still allowed by $B({\bar B}_s \to \tau^+ \tau^-)$ and
$B({\bar B} \to X_s\, \tau^+\, \tau^-)$.

Although the $c$-quark loop contribution with both couplings from NP
cannot be large enough to compete with the SM contribution,
there can be an interference term between the SM and the NP
in this case.
This term is given by
\bea
&&\Ga_{12}^{s,Z'+{\rm SM}} = -\frac{ G_F^2 m_b^2 \la_c}{ \pi} 
\l(m_Z^2 \over m_{Z'}^2 \r) \sqrt{1-4 x_c} \nl
&\times& \Bigg[ 4 B_{sb}^L B_{cc}^L \Big\{ (1-x_c)  \ME{O_{LL}} 
+  (1+2 x_c) \ME{\wt{O}_{LR}} \Big\} 
+  6 B_{sb}^L B_{cc}^R  x_c \ME{O_{LL}} 
+  12 B_{sb}^R B_{cc}^R  \ME{O_{LR}}  \nl
&+& 4 B_{sb}^R B_{cc}^L \Big\{ (1-x_c)  \ME{O_{LR}} 
+ (1+2 x_c)   \ME{\wt{O}_{LR}} \Big\} \Bigg].
\label{eq:G12_int}
\eea
Here again Eq.~(\ref{eq:G12_int}) can be comparable with $\Ga_{12}^{s,{\rm SM}}$
when $B_{sb}^i B_{cc}^j \sim {\cal O}(1)$ $(i,j=L,R)$.

\section{Numerical Results}
\label{sec:num_res}

\subsection{\bf\boldmath $Z$FCNC}

In the numerical study of the $Z$FCNC model, we scan the allowed
values of the couplings after imposing the constraints from
$|\De_s|\equiv \De M_s/\De M_s^{\rm SM}$ [Eq.~(\ref{DMsZFCNC})] and
$B({\bar B} \to X_s\, \mu^+ \,\mu^-)$\footnote{There can be other
  constraints, such as those from the forward-backward asymmetry in $B
  \to K^*\, l^+ \, l^-$~\cite{Bobeth:2008ij}. However, we do not
  include these in our analysis, since, as can be seen below in
  Fig.~\ref{fig:vqm-asl-dg}, the two constraints used are sufficient
  to rule out the $Z$FCNC model as an explanation of $S_{\psi\phi}$
  and $a_{\rm sl}^s$.}.  The general expression for the branching
ratio of ${\bar B} \to X_s\, l^+\, l^-$, where $l=e,\mu,\tau$, is
given by \cite{Alok:2010zd}
\beq
B({\bar B} \to X_s\, l^+\, l^-) = B_0\int_{4m^2_l/m^2_b}^{(1-m_s/m_b)^2} \Big(B_{SM}(z) + 
B_{SM{\hbox{-}}VA}(z) +  B_{VA}(z)\Big)\, dz \;.
\label{incl-br}
\eeq
The expressions for $B_0$, $B_{SM}$, $B_{SM{\hbox{-}}VA}$, and
$B_{VA}$ are given in Ref.~\cite{Alok:2010zd}. We do not repeat them
here. We note only that the integrand depends on the NP couplings
$R_V, R_A, R'_V$ and $R'_A$. In the $Z$FCNC model, the NP couplings
$R'_{V,A}=0$, and the $R_{V,A}$ are given by \cite{Mohanta:2008}
\beq
R_V = \frac{2\pi}{\alpha}\frac{U_{sb}}{V_{tb}V^*_{ts}}\left(-\frac{1}{2} + 2 \sin^2 \theta_W\right) ~~,~~~~
R_A = \frac{\pi}{\alpha}\frac{U_{sb}}{V_{tb}V^*_{ts}}\;.
\eeq

We define $q^\mu$ as the sum of the 4-momenta of the $\mu^+$ and
$\mu^-$ in ${\bar B} \to X_s\, \mu^+\, \mu^-$.  The theoretical predictions
for the branching ratio of this decay are not reliable over the whole
$q^2$ region due to the presence of charm resonances at intermediate
$q^2$ ($7$~GeV$^2 \leq q^2 \leq 12$~GeV$^2$). The predictions are
relatively more robust for low $q^2$ ($1 \,{\rm GeV^2} \leq q^2 \leq
6\, {\rm GeV^2}$) and high $q^2$ ($14.4\, {\rm GeV^2} \leq q^2 \leq
m_b^2$). However, the two regions have different sensitivities to the
short-distance physics. The dominant contribution to ${\bar B} \to X_s\,
\mu^+\, \mu^-$ in the low-$q^2$ region comes from a virtual photon,
whereas the dominant contribution in the high-$q^2$ region comes from
the $Z$ and $W$. Since we are interested specifically in the $Z{\ol
  b}s$ coupling, we use the branching ratio in the high-$q^2$ region
to constrain the $Z$FCNC parameter space.

%%%%%%%%%%%%%%%%%%%%%%%%%%%%%%%%%%%%%%%%%%%%
\FIGURE{
\centering
\begin{tabular}{c}
\epsfig{file=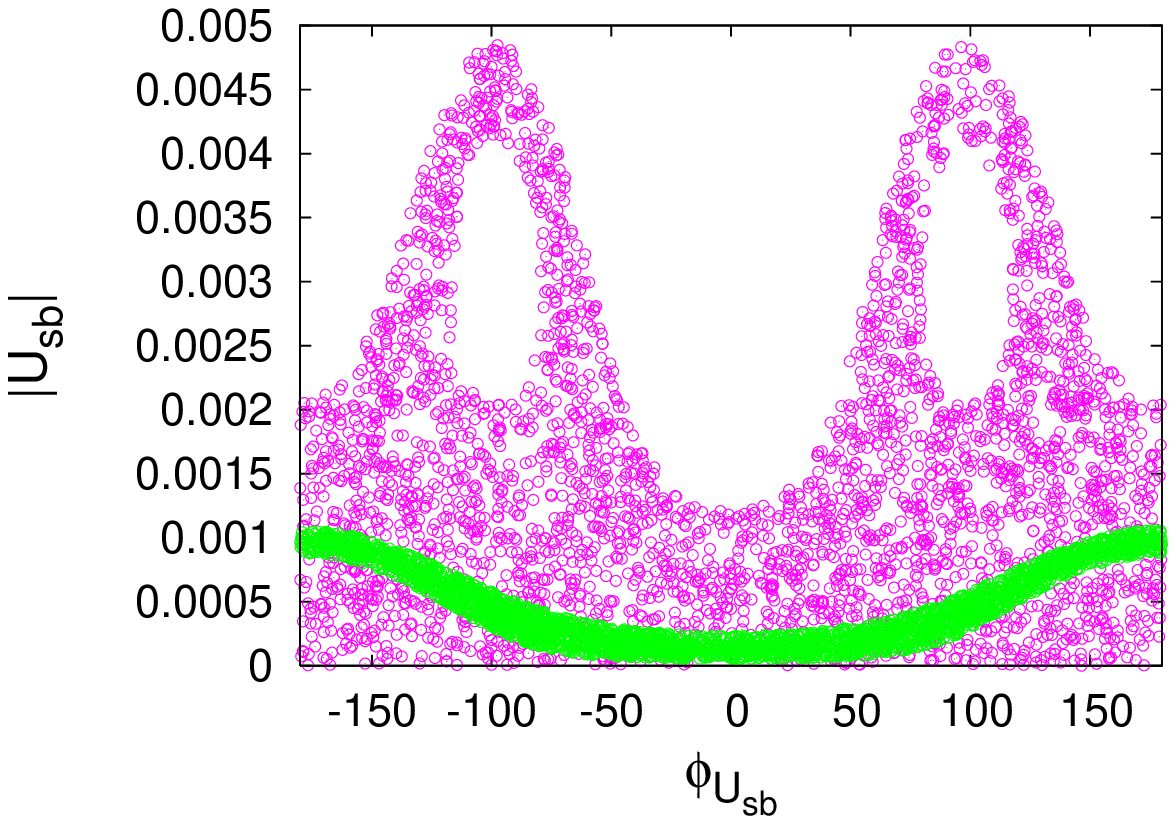,width=0.6\linewidth,clip=}
\end{tabular}
\caption{ The $\phi_{U_{sb}}$-$|U_{sb}|$ parameter space allowed by
  $|\De_s|$ (magenta) and $B({\bar B} \to X_s\, \mu^+ \,\mu^-)$ (green).
\label{fig:vqm-dm}}}
%%%%%%%%%%%%%%%%%%%%%%%%%%%%%%%%%%%%%%%%%%%%

As described earlier, $U_{sb}$ denotes the $Z{\bar b}s$ coupling
generated in the $Z$FCNC model [Eq.~(\ref{Usb})]. The parameters of
the model are therefore the magnitude and the phase of this coupling,
$|U_{sb}|$ and $\phi_{U_{sb}} \equiv \arg
U_{sb}$. Fig.~\ref{fig:vqm-dm} shows the allowed region in
($\phi_{U_{sb}}$, $|U_{sb}|$) space due to the measurements of
$|\De_s|$ (magenta) and $B({\bar B} \to X_s\, \mu^+\, \mu^-)$ (green).  Using
constraints only from $|\De_s|$, we see that the full range of
$\phi_{U_{sb}}$ is obtained only when $|U_{sb}| \lesssim 0.001$,
while $|U_{sb}|$ can be as large as 0.005 in the constrained regions
of $\phi_{U_{sb}}$. However, if we include the constraints from $B({\bar B}
\to X_s\, \mu^+\, \mu^-)$, the ($\phi_{U_{sb}}$, $|U_{sb}|$) space is
reduced considerably. From Fig.~\ref{fig:vqm-dm}, one sees that $B({\bar B}
\to X_s\, \mu^+\, \mu^-)$ limits the upper value of $|U_{sb}|$ to be
around 0.001.

Using these allowed couplings, we can calculate the observables
$S_{\psi\phi}$, $a_{\rm sl}^s$ and $\De \Ga_s$, and compare these
values with the measurements. The 1$\sigma$ ranges of these
observables are
\bea
S_{\psi\phi} & : & [-0.91 , -0.47] ~, \nl
a_{\rm sl}^s & : & [-17.7 , -7.7 ] \times 10^{-3} ~, \nl
\De \Ga_s & : & \pm [ 0.084 , 0.208 ] \; {\rm ps}^{-1} ~. 
\eea
In addition, we can examine correlations between the observables.

%%%%%%%%%%%%%%%%%%%%%%%%%%%%%%%%%%%%%%%%%%%%
\FIGURE{
\centering
\begin{tabular}{cc}
\epsfig{file=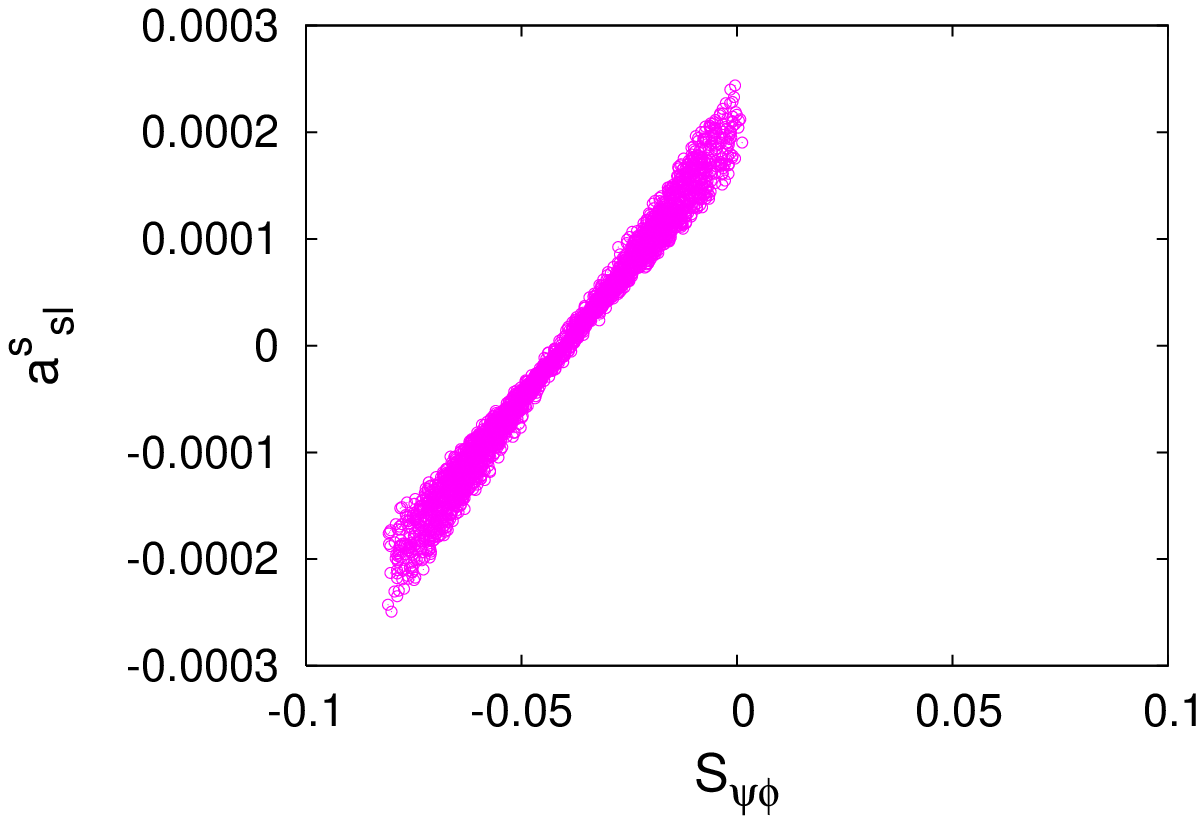,width=0.48\linewidth,clip=}
\epsfig{file=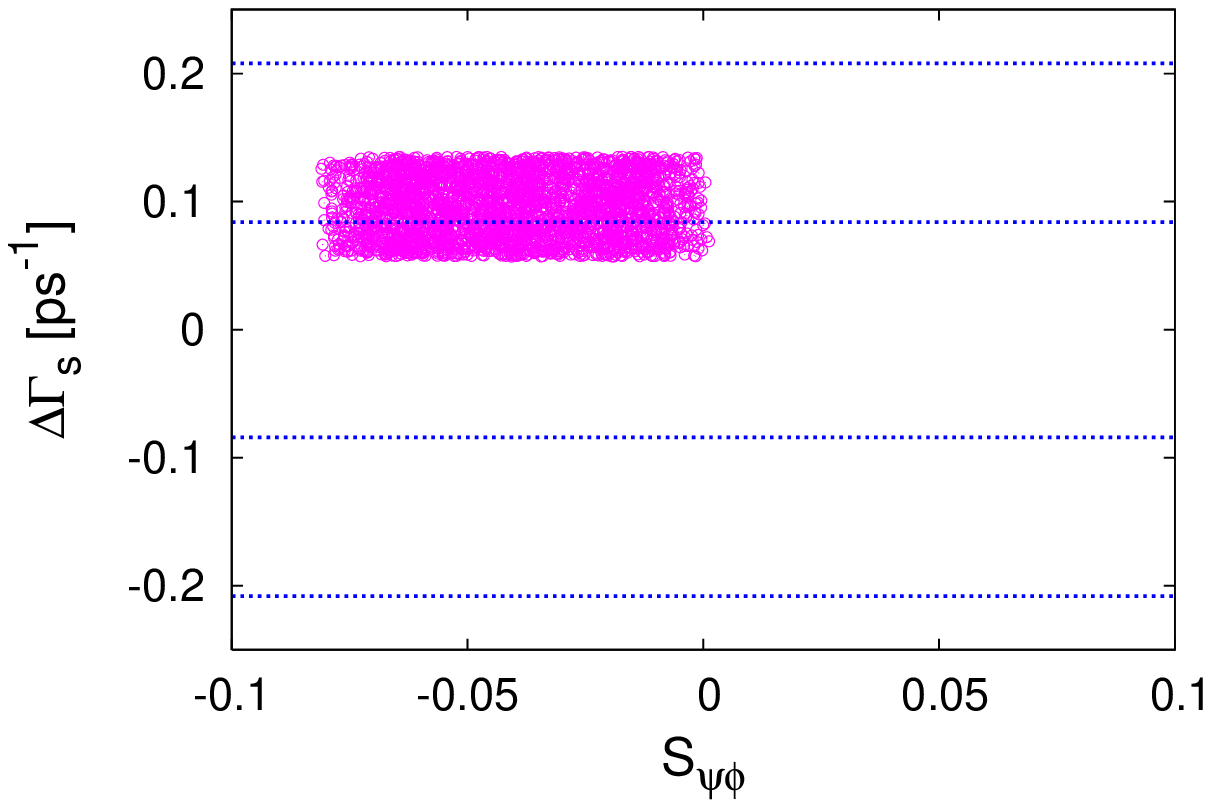,width=0.48\linewidth,clip=}
\end{tabular}
\caption{The $a^s_{sl}$-$S_{\psi \phi}$ correlation plot (left panel)
  and the $\Delta \Gamma_s$-$S_{\psi \phi}$ correlation plot (right
  panel). The horizontal lines in the right panel indicate the
  1$\sigma$ experimental allowed range for $\Delta \Gamma_s$.
\label{fig:vqm-asl-dg}}}
%%%%%%%%%%%%%%%%%%%%%%%%%%%%%%%%%%%%%%%%%%%%

As a consequence of the strong constraint on the size and phase of
$U_{sb}$, we see in Fig.~\ref{fig:vqm-asl-dg} that it is not possible
to explain the present 1$\sigma$ ranges of $S_{\psi\phi}$ and $a_{\rm
  sl}^s$ in the $Z$FCNC model. Still, this model can produce values
for these observables which are closer to the experimental data. In
particular, one can have $a_{\rm sl}^s \sim -3 \times 10^{-4}$, which is
about 10 times larger in magnitude than the SM prediction of $\sim 2
\times 10^{-5}$. And $S_{\psi\phi}$ can be decreased to $-0.08$, as
compared to the SM value of $-0.04$.

\subsection{\bf\boldmath $Z'$FCNC}

As discussed earlier, a large NP contribution to $\Ga_{12}^{s,{\rm
    NP}}$, and hence to $a^s_{sl}$, may come from the NP $b \to
s\tau^+ \tau^-$ and/or $b \to s c \bar{c}$ operators. Since the two
operators need not be related to one another, a large NP effect on $b
\to s \tau^+ \tau^-$ but not on $b \to s c \bar{c}$, or vice versa, is
conceivable. Thus, in analyzing the $Z'$FCNC model, we examine whether
it is possible to explain the present 1$\sigma$ measurement of
$a^s_{sl}$ if only one of the NP operators is present.

\subsubsection{\bf\boldmath NP $b \to s \tau^+ \tau^-$ operator only}

We first consider the case where the NP contribution to
$\Ga_{12}^{s,{\rm NP}}$ comes only from the $b \to s \tau^+
\tau^-$ operator. In our numerical study we scan the allowed values of
the $Z'$FCNC couplings imposing the constraints from $|\De_s|$, $B({\bar B}
\to X_s\, \tau^+ \,\tau^-)<5\%$ and $B({\bar B}_s \to \tau^+ \,\tau^-)<5\%$.
$|\De_s|$ can be obtained using Eqs.~(\ref{effH-M12s}) and
(\ref{ME-M12s}), and $B({\bar B} \to X_s\, \tau^+ \,\tau^-)$ is given by
Eq.~(\ref{incl-br}) with $l=\tau$. The branching ratio of ${\bar B}_s \to
\tau^+ \,\tau^-$ is \cite{Alok:2010zd}
\begin{equation}
B({\bar B}_s \to \tau^+ \,\tau^-) = \frac{G^2_F \alpha^2
m^5_{B_s} f_{B_s}^2 \tau_{B_s}}{64 \pi^3}
|V_{tb}^{}V_{ts}^{\ast}|^2 \sqrt{1 - \frac{4 m_\tau^2}{m_{B_s}^2}}\;
\Bigg| \frac{2 m_\tau}{m^2_{B_s}} (C_{10}+R_A-R'_A)\Bigg|^2\;.
\label{bmumu-BR}
\end{equation}
In the $Z'$FCNC model, the NP couplings are given by
\bea
R_V &=& -\frac{4\pi}{\alpha}\frac{m^2_Z/m^2_{Z'}}{V_{tb}V^*_{ts}} B^L_{sb}(B^L_{\tau\tau}+B^R_{\tau \tau})\;, \nl
R_A &=& \frac{4\pi}{\alpha}\frac{m^2_Z/m^2_{Z'}}{V_{tb}V^*_{ts}} B^L_{sb}(B^L_{\tau\tau}-B^R_{\tau \tau})\;,\nl
R'_V &=& -\frac{4\pi}{\alpha}\frac{m^2_Z/m^2_{Z'}}{V_{tb}V^*_{ts}} B^R_{sb}(B^L_{\tau\tau}+B^R_{\tau \tau})\;, \nl
R'_A &=& \frac{4\pi}{\alpha}\frac{m^2_Z/m^2_{Z'}}{V_{tb}V^*_{ts}} B^R_{sb}(B^L_{\tau\tau}-B^R_{\tau \tau})\;.
\eea
These couplings are constrained by the above observables.  However,
the key point is that, because the bounds on the NP $b \to s \tau^+
\tau^-$ transition are still weak, the constraints are not severe.

%%%%%%%%%%%%%%%%%%%%%%%%%%%%%%%%%%%%%%%%%%%%
\FIGURE{
\centering
\begin{tabular}{cc}
\epsfig{file=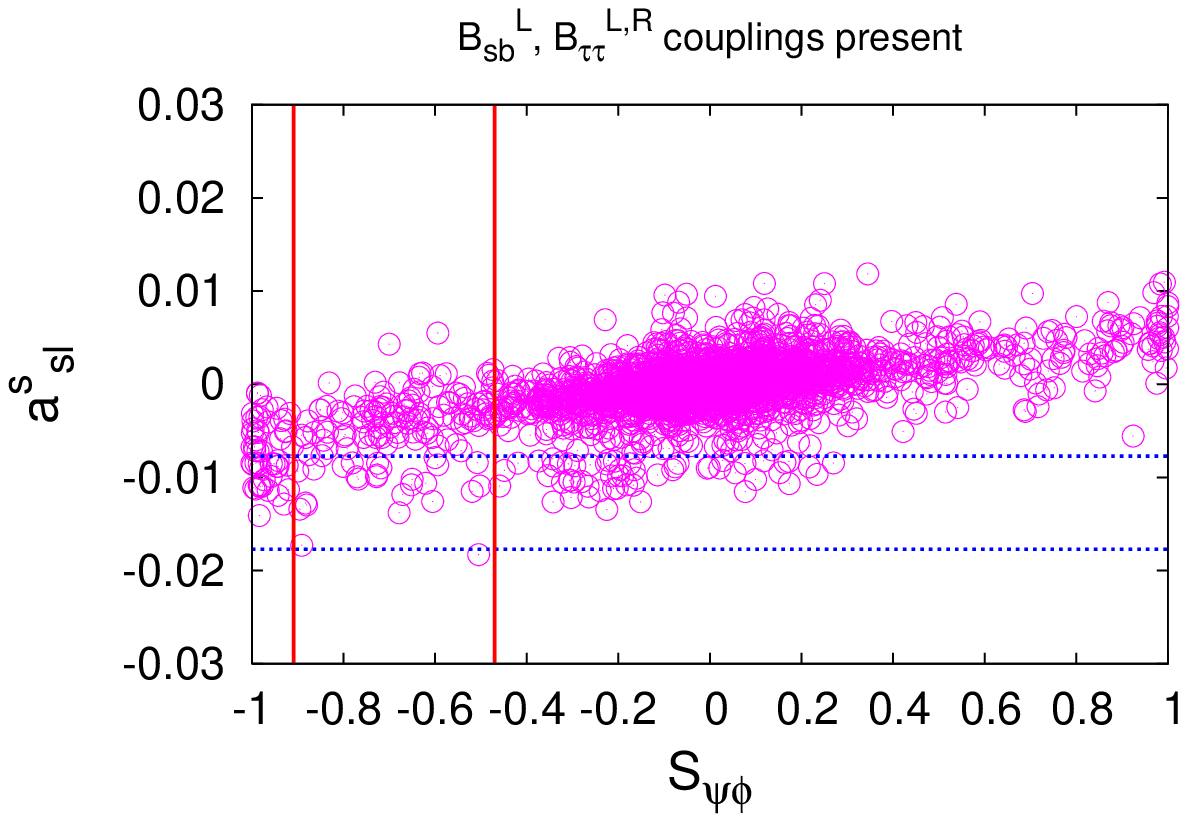,width=0.48\linewidth,clip=}
\epsfig{file=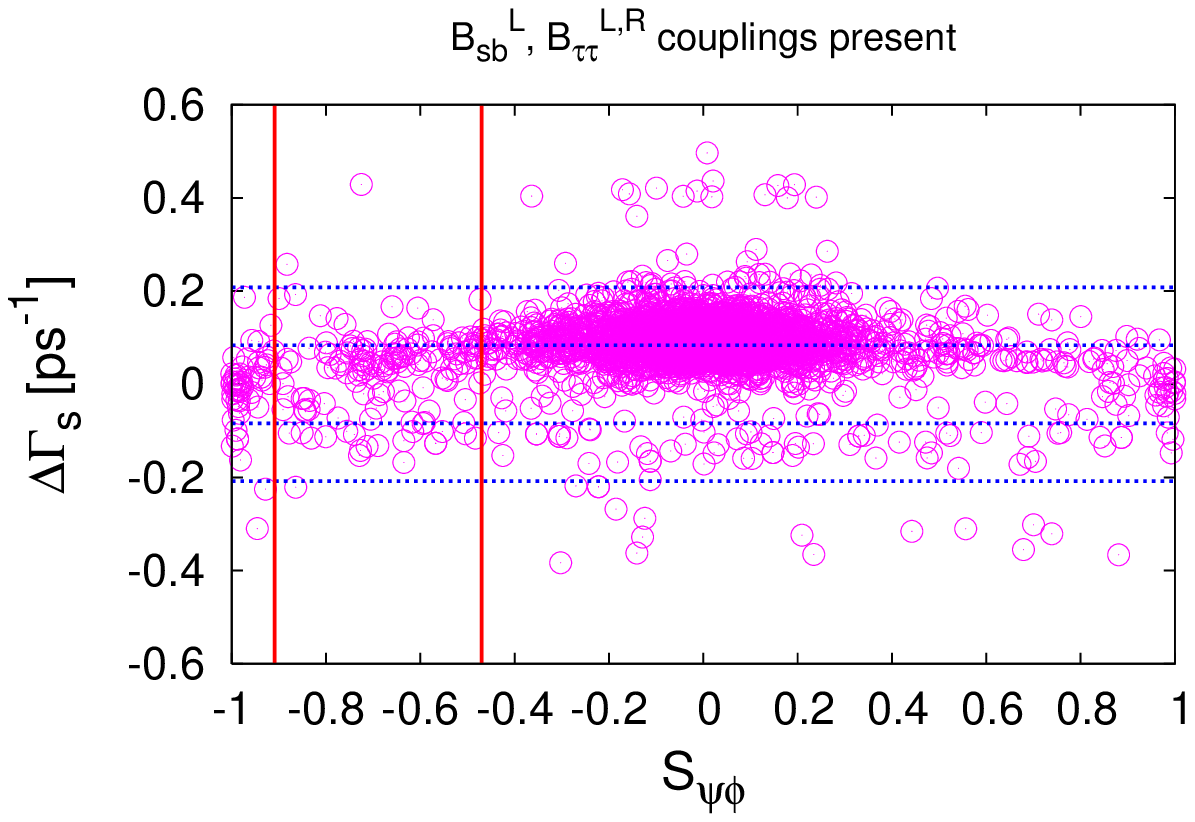,width=0.48\linewidth,clip=}
\end{tabular}
\caption{The $a^s_{sl}$-$S_{\psi \phi}$ correlation plot (left panel)
  and the $\Delta \Gamma_s$-$S_{\psi \phi}$ correlation plot (right
  panel), for the case where only $B_{sb}^L$ and $B_{\tau \tau}^{L,R}$
  couplings are present.  The horizontal and vertical lines indicate the
  1$\sigma$ experimental allowed ranges for the observables.}
\label{fig:bttL}
}
%%%%%%%%%%%%%%%%%%%%%%%%%%%%%%%%%%%%%%%%%%%%

Fig.~\ref{fig:bttL} shows the $a^s_{sl}$-$S_{\psi \phi}$ and $\Delta
\Gamma_s$-$S_{\psi \phi}$ correlation plots for the case where only
$B_{sb}^L$ and $B_{\tau \tau}^{L,R}$ couplings are present. We make
the following observations:
\begin{itemize}

\item $S_{\psi \phi}$ can take any value between $-1$ and $1$.

\item The $Z'$FCNC model gives values for $a^s_{sl}$ within its
  1$\sigma$ range.

\item One can simultaneously reproduce the present measurements of all
  three observables, $S_{\psi \phi}$, $a^s_{sl}$ and $\Delta
  \Gamma_s$, within their 1$\sigma$ ranges.

\end{itemize}
Obviously, one obtains the same conclusions if a nonzero $B_{sb}^R$ is
also present.

\subsubsection{\bf\boldmath NP $b \to s c \bar{c}$ operator only} 

We now consider the case where the NP contribution to
$\Ga_{12}^{s,{\rm NP}}$ comes only from the $b \to s c \bar{c}$
operator. In our numerical study we scan the allowed values of the
$Z'$FCNC couplings imposing the constraints from $|\De_s|$ and the
indirect CP asymmetry in ${\bar B}_d \to J/\psi K_S$. This latter
constraint arises as follows. In the presence of NP in the decay $b
\to s c \bar{c}$, the effective measured $\sin 2\beta$ in ${\bar B}_d
\to J/\psi K_S$ is given by \cite{psiphigeneral}
\beq
\sin 2\beta^{meas} = \sin2\beta + 2 |r| \cos2\beta
 \sin\varphi\cos\delta^a ~.
\label{Spsiks}
\eeq
Here, the NP parameters are defined similarly to those in
Eq.~(\ref{Spsiphi}): $|r|$ is the ratio of magnitudes of the NP and SM
$b \to s c \bar{c}$ amplitudes, $\varphi$ is the NP weak phase, and
$\delta^a$ is the SM strong phase.  The true value of $\sin 2\beta$ is
taken from the fit to the sides of the unitarity triangle: $\sin
2\beta=0.731\pm0.038$, while the experimental measurement gives $\sin
2\beta^{meas}=0.668\pm0.028$ \cite{Charles:2004jd}.  Using these
values, $|r|$ is estimated to be \cite{psiphigeneral}
\beq
|r|=(4.6\pm 3.5)\%\,.
\eeq
Thus, at 1$\sigma$, $|r|\leq 8.1\%$ is permitted.  This upper bound on
$|r|$, along with the measurement of $|\De_s|$, constrain the NP
couplings.

%%%%%%%%%%%%%%%%%%%%%%%%%%%%%%%%%%%%%%%%%%%%
\FIGURE{
\centering
\begin{tabular}{cc}
\epsfig{file=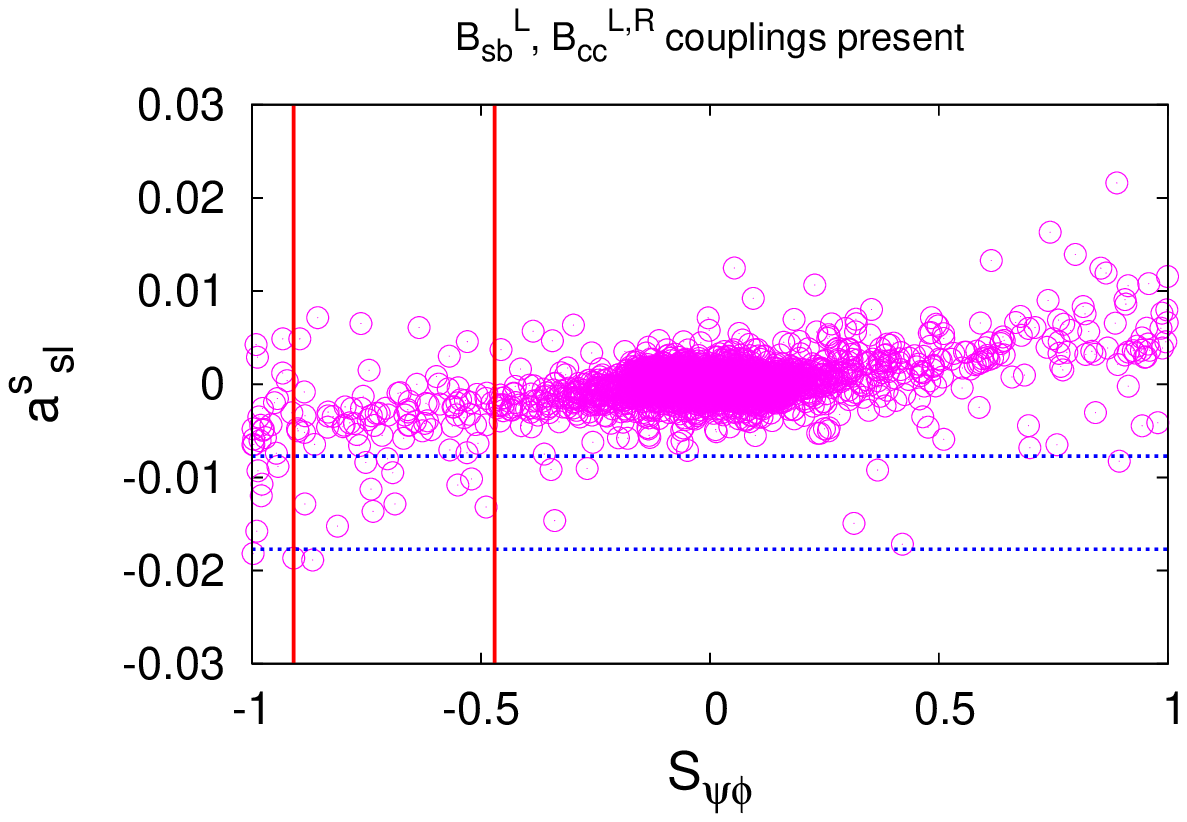,width=0.48\linewidth,clip=}
\epsfig{file=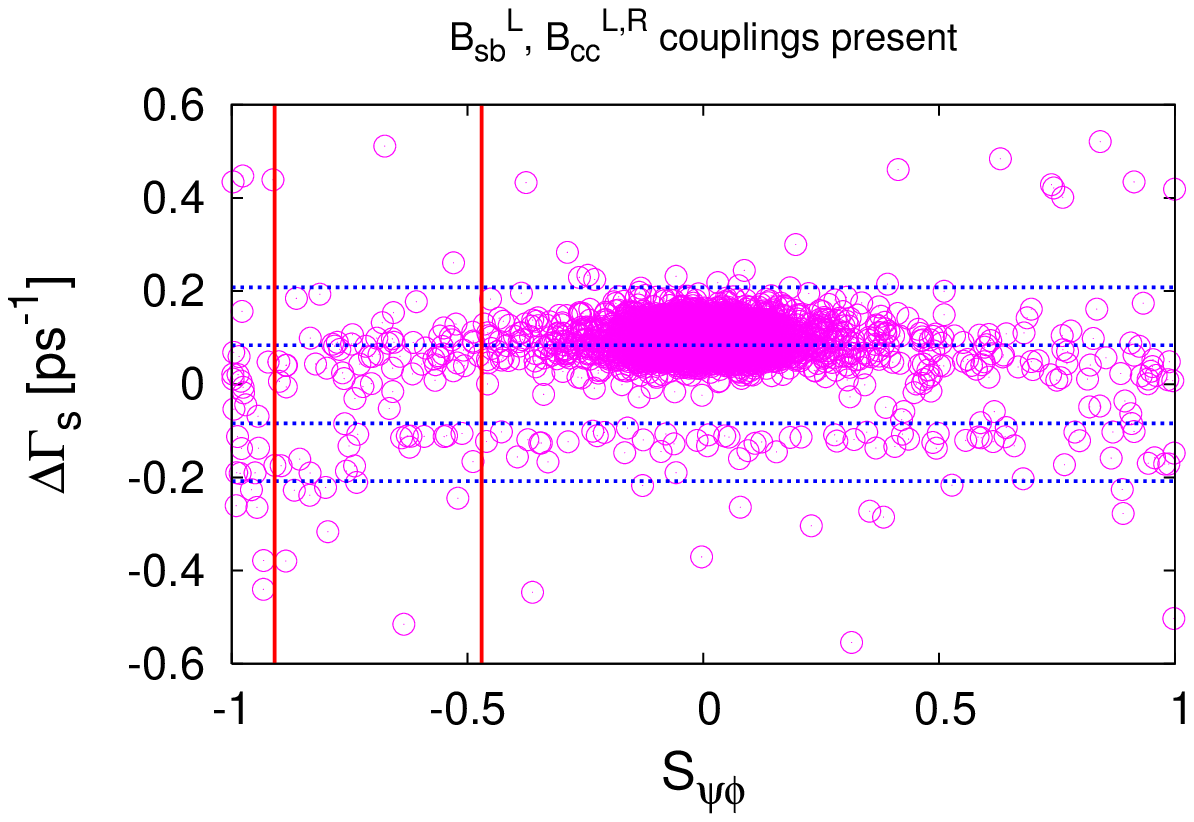,width=0.48\linewidth,clip=}
\end{tabular}
\caption{The $a^s_{sl}$-$S_{\psi \phi}$ correlation plot (left panel)
  and the $\Delta \Gamma_s$-$S_{\psi \phi}$ correlation plot (right
  panel), for the case where only $B_{sb}^L$ and $B_{cc}^{L,R}$
  couplings are present.  The horizontal and vertical lines indicate
  the 1$\sigma$ experimental allowed ranges for the observables.
\label{fig:bccL}}
}
%%%%%%%%%%%%%%%%%%%%%%%%%%%%%%%%%%%%%%%%%%%%

Fig.~\ref{fig:bccL} shows the $a^s_{sl}$-$S_{\psi \phi}$ and $\Delta
\Gamma_s$-$S_{\psi \phi}$ correlation plots for the case where only
$B_{sb}^L$ and $B_{cc}^{L,R}$ couplings are present. The results are
the same as for the case where a NP $b \to s \tau^+ \tau^-$
operator was added:
\begin{itemize}

\item $S_{\psi \phi}$ can take any value between $-1$ and $1$.

\item The $Z'$FCNC model gives values for $a^s_{sl}$ within its
  1$\sigma$ range.

\item One can simultaneously reproduce the present measurements of all
  three observables, $S_{\psi \phi}$, $a^s_{sl}$ and $\Delta
  \Gamma_s$, within their 1$\sigma$ ranges.

\end{itemize}

We therefore see that the $Z'$FCNC model can reproduce the present
measurements of $S_{\psi\phi},\, \De \Ga_s$ and $a_{\rm sl}^s$ within
their 1$\sigma$ ranges if the NP contribution comes only from either
the $b \to s\tau^+ \tau^-$ or $b \to s c \bar{c}$ operators.

\section{Conclusions}
\label{sec:concl}

Recently, the D\O\ Collaboration measured the CP-violating like-sign
dimuon charge asymmetry in neutral $B$ decays and found a 3.2$\sigma$
difference from the prediction of the standard model (SM). Combining
the D\O\ result with other measurements, we obtain the semileptonic
charge asymmetry for $B_s$ mesons, $(a_{\rm sl}^s)_{\rm ave} = -(12.7
\pm 5.0) \times 10^{-3}$, which is still $\sim 2.5\sigma$ away from
the SM prediction.

A non-SM $a_{\rm sl}^s$ can be explained theoretically by a
new-physics (NP) contribution to $B_s$-$\ol{B}_s$ mixing, specifically
to $M_{12}^s$ and $\Ga_{12}^s$, the dispersive and absorptive parts of
the mixing. It is usually assumed that $\Ga_{12}^s$ is dominated by
the SM tree-level $b \to s c \bar{c}$ coupling, and that it is not
affected by the NP. Instead, NP is considered only in $M_{12}^s$.
However, in this case, the measured value of $(a_{\rm sl}^s)_{\rm
  ave}$ within its 1$\si$ range cannot be explained. Thus, NP in
$\Ga_{12}^s$ is necessary.

There are some NP models (leptoquarks, R-parity-violating SUSY) whose
contributions to $\Ga_{12}^s$ can compete with the SM. In this paper,
we examine models with flavor-changing neutral gauge bosons ($Z$ or
$Z'$) to see if they can explain $(a_{\rm sl}^s)_{\rm ave}$. The
models must also reproduce the measured values of the indirect CP
asymmetry $S_{\psi\phi}$ in ${\bar B}_s \to J/\psi \phi$ and $\De \Ga_s$.

In our models we assume that the main contribution to $\Ga_{12}^s$
comes from diagrams involving $\tau$ and/or $c$ loops. The point is
that the current constraints on NP $b \to s f \ol{f}$ ($f=\tau$, $c$)
transitions are quite weak. We find that the model with
flavor-changing $Z$ couplings ($Z$FCNC) cannot reproduce the present
measured values of $S_{\psi\phi}$ and $a_{\rm sl}^s$ within their
1$\sigma$ ranges. Still, the $Z$FCNC model can lead to modified values
for $S_{\psi\phi}$ and $a_{\rm sl}^s$ which are closer to the
experimental data than the SM predictions.

On the other hand, in the model with flavor-changing $Z'$ couplings
($Z'$FCNC), even after imposing the current experimental constraints,
the present measurements of $S_{\psi \phi}$, $a^s_{sl}$ and $\Delta
\Gamma_s$, can all be simultaneously reproduced within their 1$\sigma$
ranges. Indeed, the full range of $S_{\psi\phi}$ is allowed. We
therefore see that the $Z'$FCNC model is another type of NP which can
explain the measured value of the D\O\ dimuon charge asymmetry in the
$B$ system.

\bigskip
\noindent
{\bf Acknowledgments}: We thank F. J. Botella for very helpful
comments about the $Z$FCNC model.  This work was financially supported
by NSERC of Canada (AKA,DL).  SB is grateful for the hospitality of
the Universit\'e de Montr\'eal, where part of this work was done. SB
also acknowledges financial support from the Basic Science Research
Program through the National Research Foundation of Korea (NRF),
funded by the Ministry of Education, Science and Technology
No. 20100028004.

\end{document}